\documentclass[]{spie}  

\usepackage{amsmath,amsfonts,amssymb}
\usepackage{graphicx}
\usepackage[colorlinks=true, allcolors=blue]{hyperref}

\title{The ViSta method for optimized stacking of broadband interferometric data in the Fourier domain}

\author[a]{Martina Torsello}
\author[b,a]{Marcella Massardi}
\author[b]{Elisabetta Liuzzo}
\author[a,c]{Gayathri Gururajan}
\author[a]{Francesca Perrotta}
\author[a,d,c,e]{Andrea Lapi}

\affil[a]{Scuola Internazionale Superiore di Studi Avanzati, Via Bonomea 265, 34136 Trieste, Italy}
\affil[b]{INAF - Istituto di Radioastronomia - Italian ALMA Regional Centre, Via Gobetti 101, 40129 Bologna, Italy}
\affil[c]{Institute for Fundamental Physics of the Universe (IFPU), Via Beirut 2, 34014 Trieste, Italy}
\affil[d]{INAF - Istituto di Radioastronomia, Via Gobetti 101, 40129 Bologna, Italy}
\affil[e]{Istituto Nazionale Fisica Nucleare (INFN), Sezione di Trieste, Via Valerio 2, 34127 Trieste, Italy}

\authorinfo{Send correspondence to M.T., E-mail: mtorsell@sissa.it, E-mail: martina.torsello@gmail.com}

\begin{document} 
\maketitle
\let\oldthefootnote\thefootnote
\let\thefootnote\relax
\footnotetext{%
  \textcopyright\ 2026 Society of Photo-Optical Instrumentation Engineers (SPIE).
  Cite as: M.~Torsello, M.~Massardi, E.~Liuzzo, G.~Gururajan, F.~Perrotta,
  and A.~Lapi, ``The ViSta method for optimized stacking of broadband
  interferometric data in the Fourier domain,'' Proc.\ SPIE \textbf{14153},
  Radio Telescopes, Technologies, and Methods, 141531X (2026),
  \url{https://dx.doi.org/10.1117/12.3103792}.
}
\let\thefootnote\oldthefootnote

\begin{abstract}
We present the optimized version of \texttt{ViSta}, a visibility-domain stacking method that combines interferometric observations in the Fourier domain from radio to sub-millimeter wavelengths. By stacking visibilities directly and transforming them into the rest frame, \texttt{ViSta} enhances the signal, suppresses noise, and improves image reconstruction through extended \textit{uv}-coverage. \texttt{ViSta} outperforms image stacking when individual sources are too faint to detect, achieving higher SNR in the low-signal and extended regime. This new optimized version features a C++/OpenMP kernel which replaces CASA demanding functions, enabling GPU acceleration while minimising memory usage and intermediate data products. The method is highly flexible, allowing stacking regardless of array configuration, spectral setup, or telescope. In the SKA era, where visibilities will not be routinely preserved, \texttt{ViSta} helps in assessing the information lost in the image-plane transition and in exploiting the vast amount of data still stored in interferometric archives.

\end{abstract}


\section{Introduction}

At radio and sub-millimetre wavelengths, galaxies reveal their physical complexity unobscured by dust, offering a unique window into the processes that govern their evolution across cosmic time. Through a rich spectrum of molecular and atomic diagnostic lines, these wavelengths encode information on the cold molecular gas reservoirs that fuel star formation, the warm ionised phases shaped by stellar winds, supernovae, and AGN-driven outflows, and the dust continuum that records the growth history of stellar mass distribution. This spectral window becomes uniquely powerful at high redshift ($z\sim2$--3), where the peak of the cosmic star formation rate density and the rapid assembly of the most massive galaxies are redshifted precisely into the millimetre and sub-millimetre bands. Detecting and characterising such emission at these wavelengths is beyond the reach of any single dish. By correlating signals between antenna pairs spread over baselines of up to tens of kilometers, modern radio and sub-millimetre interferometers synthesise an effective aperture far larger than any single telescope could achieve, reaching the angular resolutions and sensitivities required to study the ISM of high-redshift galaxies across cosmic time.
However, unlike optical telescopes which directly record a focused image on a detector, interferometers measure the sky indirectly. Each antenna pair produces a correlated signal, called visibility, that samples a single Fourier component of the sky brightness distribution. Each visibility corresponds to a point in the two-dimensional Fourier conjugate of the sky, i.e. the \textit{uv}-plane, whose coordinates are set by the projection of the baseline between the two antennas onto the plane of the sky. Reconstructing an image ready for scientific analysis therefore requires inverting this incomplete Fourier sampling through computationally demanding and inherently lossy operations such as gridding, deconvolution, and imaging.

The major interferometric facilities (ALMA, the VLA, MeerKAT and others) preserve the raw visibility data alongside the final data products, storing them in standardised archives open to the public. The full Fourier-domain information content of every observation is retained here and remains available for re-analysis, combination with other datasets or exploitation for science cases beyond the original proposal. The resulting archives are therefore vast repositories of largely unexploited information. For example, the ALMA Science Archive (ASA) alone accumulates 300--400\,TB of raw and reduced data per year \cite{Stoehr2024}, and as of 2024 archival data contributed to 40\% of all ALMA-related publications \cite{Stoehr2026}. After three years from data delivery, only $\sim$30\% of projects from recent cycles have resulted in a publication by the PI group, with a median time from data delivery to first publication of 2.1 years \cite{Stoehr2026}. Moreover, even when a dataset is published, only a fraction of its spectral content is typically analysed. Most observations focus on the brightest accessible features, while the bulk of this diagnostic information remains below the detection threshold of any individual pointing. When the signal of interest is too faint to be detected with the sensitivity of an individual observation, but a statistical sample of similar sources is available, a natural solution is to combine them through stacking techniques. By co-adding $N$ observations, the noise averages down as $1/\sqrt{N}$, while the common signal accumulates, making it possible to recover average properties of the population that would be entirely inaccessible in any single pointing. 
Stacking is most commonly performed in the image domain, where it is straightforward to align images from different observations. However, stacking directly in the \textit{uv}-plane has been shown to be intrinsically more efficient: it preserves the Gaussian statistics of the noise, avoids the artifacts introduced by incomplete \textit{uv}-coverage in the individual images, and requires only a single deconvolution of the final stacked product \cite{Hancock2011, Lindroos2015, Hill2024}. 
These techniques, however, were developed for sources at the same redshift observed with the same spectral setup, so that their visibilities already lie on a coherent \textit{uv}-plane and can be directly combined. Extending this approach to heterogeneous samples (sources at different redshifts, observed with different array configurations, frequency setups, and telescopes) requires mapping each dataset onto a common rest-frame \textit{uv}-plane before any combination is possible.

To achieve this purpose, we developed \texttt{ViSta} \cite{Torsello2025}, a method that rescales, recentres, and regrids each input dataset onto a common rest-frame \textit{uv}-plane before combining them into a single unified observation. In Ref. \citenum{Torsello2025}, we validated the method on both simulated and real ALMA data, demonstrating that visibility stacking reliably recovers the flux of sources hidden in the noise across a range of source morphologies and array configurations, and that its advantage over image-plane stacking grows progressively as the individual source sensitivity decreases. Stacking in the \textit{uv}-plane eliminates the need for $N$ individual deconvolutions, which is the primary computational bottleneck of image-plane stacking. However, it introduces its own computational challenge, since manipulate the raw visibility data for a large number of observations is inherently memory and I/O-intensive. If not handled efficiently, the computational overhead of processing large numbers of raw visibility datasets risks negating the very advantage that motivates working in the visibility domain. The original implementation of \texttt{ViSta} was built on top of CASA \cite{CASATeam2022}, the standard software package for interferometric data reduction, which stores data in Measurement Set (MS) format. While well-suited for single-dataset reduction, CASA was not designed for the kind of large-scale multi-dataset processing and transformation that visibility stacking requires. Its serial execution model becomes excessively time-consuming and both memory and I/O bounded as the number of datasets grows.

In this paper we present an HPC-optimised version of \texttt{ViSta}, which replaces the CASA backend with a self-contained Python package backed by a compiled C++/OpenMP kernel with optional GPU acceleration. This new implementation removes the bottleneck described before, making \texttt{ViSta} practical as a general purpose tool for the systematic exploitation of interferometric archives at scale.

The paper is structured as follows: Section~\ref{sec:vista} summarises the \texttt{ViSta} methodology and the old code workflow; Section~\ref{sec:pipeline} describes the new pipeline architecture and the C++/GPU kernel; Section~\ref{sec:benchmarks} presents the performance benchmarks; Section~\ref{sec:conclusions} discusses the implications of this work and future developments.

\section{The ViSta method}\label{sec:vista}
This new stacking methodology was introduced and validated in \citenum{Torsello2025}, where we provided a full description of the approach and its application to ALMA archival data. The central insight of \texttt{ViSta} is that, by reprojecting all visibilities onto a coherent grid, i.e. a common rest-frame \textit{uv}-plane, observations taken at different redshifts, frequencies, and array configurations can be combined in the Fourier domain as if they were independent samplings of the same source at z$=0$. We carry out this reprojection primarily by applying a $z$-dependent rescaling of factor $(1+z)$ to all frequency dependent quantities and its inverse to the baseline vectors and \textit{uv}-coordinates of each dataset. Then, the phase center of each dataset is shifted to the source position and set to zero, while their spectral axes are regridded onto a common grid with the widest spectral coverage in the sample.  Once these steps are completed, all datasets can be concatenated into a single unified \textit{uv}-plane with significantly improved coverage and sensitivity. The resulting stacked observation is now ready for scientific analysis and we can directly produced the final image or apply further post-processing operations, depending on the specific scientific goals. This method has been validated on both simulated and real ALMA observations of resolved and unresolved sources, systematically comparing its outputs with those obtained from conventional image-plane stacking.
The main result is that \texttt{ViSta} not only reliably recovers the flux of sources hidden in the noise across all tested configurations, but its advantage over conventional image-plane stacking grows progressively as individual sensitivity decreases. As expected, working directly in the Fourier domain rather than in the image plane allows us to preserve the Gaussian nature of the noise, maintain consistency across angular scales and mitigate hole interpolation problems.
\subsection{CASA-based Implementation and its Limitations}
\begin{figure}[ht]
\centering
\includegraphics[width=\textwidth,trim={0 6cm 11cm 12cm}]{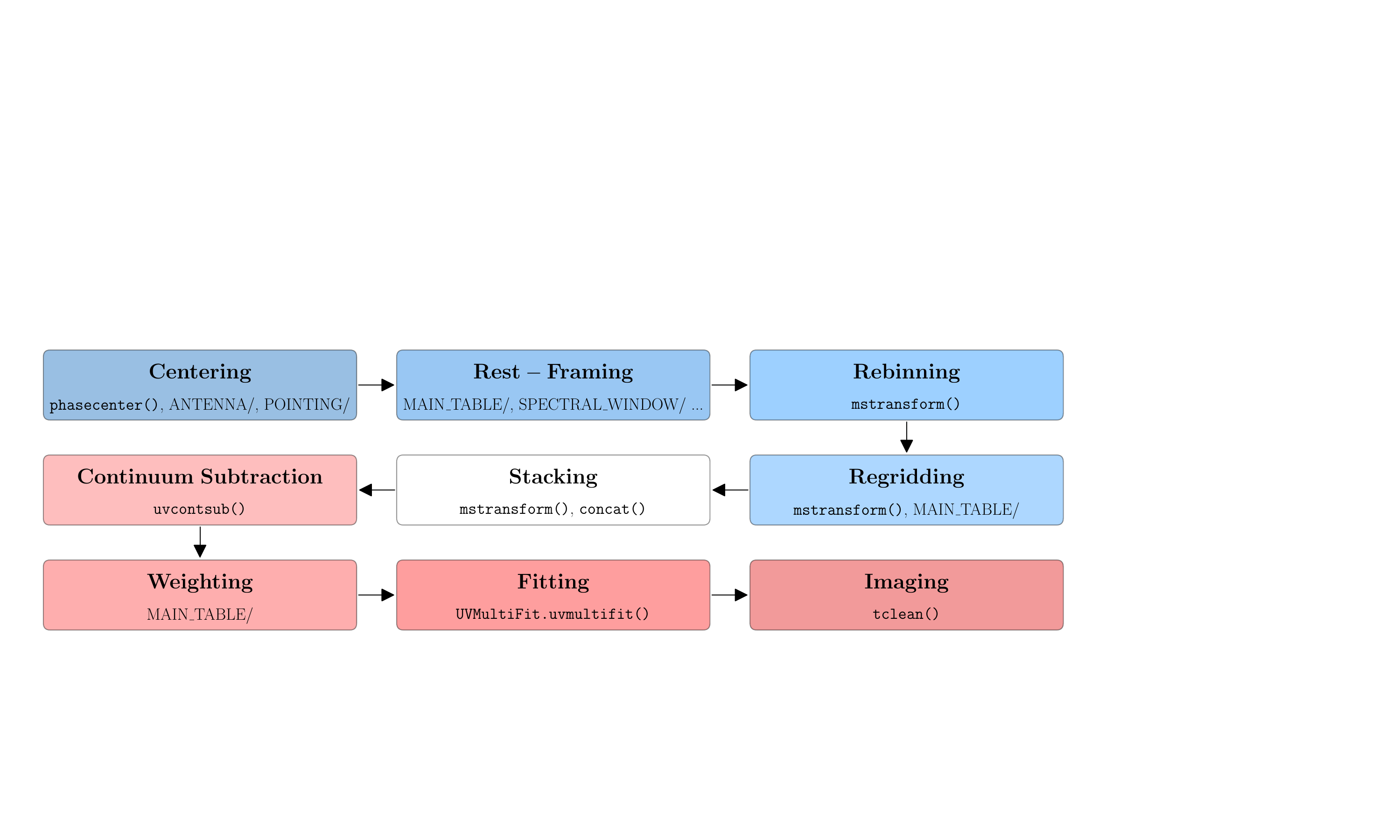}
\caption{Schematic of the original CASA-based \texttt{ViSta} pipeline \cite{Torsello2025}. Each cell represents a step of the workflow with a list of the CASA functions used and the tables that have been manually modified. The
blue cells refers to the pre-proccesing step before the stacking, while the red color cells
represents the additional post-processing operation}
\label{fig:flowchart_casa}
\end{figure}

The original \texttt{ViSta} pipeline was built on top of the CASA (Common Astronomy Software Applications) package, the reference software for ALMA data processing and the primary framework for handling MS formatted data.
The pipeline is shown in Fig. \ref{fig:flowchart_casa} and consists of a pre-processing stage (rest-framing, centering, and spectral rebinning and regridding), the stacking step itself and a set of optional post-processing operations that can be applied to the combined dataset depending on the scientific goals. These include continuum subtraction, channel reweighting to balance the contribution of each output channel and the final analysis, which can be carried out either in the image domain via standard deconvolution with \texttt{tclean}, or directly in the visibility plane through parametric fitting with UVMultiFit \cite{MartiVidal2014}. Each major pipeline stage is implemented as a separate call to different CASA standard functions, and each one of them produces a full copy of the MS to disk with the modifications before the next stage can begin. Subproducts can be eliminated directly with a specific routine when they are not needed anymore, but this mechanism multiplies inevitably storage requirements and I/O overhead. In addition, CASA retains the complete table structure of each dataset, including calibration tables, metadata and auxiliary columns that carry no relevant information for the stacking.

Furthermore, while CASA exposes parallelisation options via MPI, these remain limited in scope and unverified in practice as explicitly acknowledged in the official CASA documentation \cite{CASATeam2022}. CASA also relies on GUI-dependent tools that are unavailable on most HPC nodes. While the modular pip-wheel distribution nominally offers a more portable alternative, it introduces fragile dependency chains that break across environment updates and incurs significant overhead. Finally, CASA itself is undergoing a complete architectural overhaul, as new tools and facilities are now being built on the CASA experience in order to better meet the demands of large-scale data processing required by new facilities and telescope upgrades.
These computational limitations must therefore be addressed, otherwise the scientific advantage of working in the visibility domain risks being lost. In the image domain, for a sample of $N$ observations, the primary bottleneck is the cleaning step applied $N$ times, an operation that is both computationally demanding and potentially destructive of faint signal. Stacking in the \textit{uv}-plane of $N$ datasets requires only a single deconvolution, but this gain is only meaningful if the overall pipeline does not exceed the time and memory requirements of a standard image-plane approach.

\section{Pipeline Optimization and HPC Reimplementation}\label{sec:pipeline}

In this paper we present the new implementation of \texttt{ViSta}, which no longer relies on the CASA software, but instead consists of a self-contained Python package backed by a compiled C++/OpenMP extension module, with the possibility of running the data processing parts on GPU. This version avoids the creation of intermediate data products, minimizes memory footprint and exposes the full parallelism available on modern HPC clusters, while also being significantly faster in serial execution.

\subsection{Architecture}
Instead of using CASA table routines to handle data reading, the code relies on \texttt{dask-ms}, a library that exposes MS tables as lazy \texttt{xarray} datasets backed by Dask arrays. Each column is represented as a chunked array on which operations are accumulated into a computation graph without performing any I/O, until a \texttt{.compute()} function call triggers the actual parallel read from disk. After a pre-processing stage that reads metadata and assembles the output subtables, the pipeline is structured around a pool of concurrent threads, each dedicated to a specific operation and communicating through bounded queues.
Before the loop starts, the code performs a sequential pass over all input datasets to compute the common output grid and assemble the static output structures. The list of MSs to be stacked is provided through an input text file, where each line specifies, in order, the path to the file, the source redshift z, its sky coordinates (R.A., Dec.), and optionally the FIELD\_ID and SPW\_ID to select within that MS.
The metadata reading step then begins. Since each of these calls is small and dominated by filesystem latency rather than by the compute step, the reads for all input MS are dispatched concurrently using a \texttt{ThreadPoolExecutor}, i.e. a pool of worker threads, scheduled by Python, that execute independent tasks concurrently and collect the results once all tasks complete. This pool is separate from the reader pool discussed in the next point, which is instead design to handle heavier data chuncks.
For each observation, the \texttt{SPECTRAL\_WINDOW}, \texttt{FIELD}, and \texttt{DATA\_DESCRIPTION} subtables are read to extract the per-channel frequency array, the phase centre direction, and the \texttt{DATA\_DESC\_ID} mapping for the selected field and spectral window. From these arrays, the code computes, for every input MS, the restframed frequency range $[\nu_\mathrm{lo}(1+z), \nu_\mathrm{hi}(1+z)]$ and the restframed channel width $\Delta\nu_\mathrm{rf} = \Delta\nu_\mathrm{obs}\cdot(1+z)$. The common output channel width is then set to $\Delta\nu_\mathrm{new} = \max_i \Delta\nu_{\mathrm{rf},i}$, and the output frequency grid for each MS is placed within its restframed range, centred on the requested \texttt{central\_freq} or shifted to the nearest edge if the ideal window does not fit.
Then, the \texttt{ANTENNA} table of the output MS is built by concatenating the antenna positions of all input MS, rescaled by $1/(1+z)$, and assigning each input MS a contiguous block of antenna IDs. The remaining fundamental output subtables (\texttt{SPECTRAL\_WINDOW}, \texttt{DATA\_DESCRIPTION}, \texttt{FIELD}, \texttt{FEED}, \texttt{POLARIZATION} and \texttt{SOURCE}) are constructed entirely in memory at this stage, with one spectral window per input MS using its previously computed output grid.
Each input MS therefore contributes its own spectral window to the output dataset, and these are kept separate rather than merged into a single SPW. Keeping the SPWs distinct can be useful for subsequent analysis, and merging them would bring no benefit either in terms of memory footprint or for the final imaging step. A unified spectral window would only be required if the stacked dataset is to be fitted directly in the visibility plane with UVMultiFit \cite{MartiVidal2014}; this, however, also requires optimising the time-averaging step of the pipeline, and is left for a future update of the code as explained in section \ref{sec:postprocess}.

When all the pre-processing steps are completed, $N_\mathbf{readers}$ threads are starting the effective pipeline. Each reader thread opens a subset of the input MSs and issues a single \texttt{dask.compute()} call per spectral window, materialising visibilities, flags, UVW coordinates, and a minimal set of scalar columns into NumPy arrays in one graph evaluation, rather than triggering computation column by column. This avoids the overhead of repeated lazy-graph construction and lets the Dask scheduler overlap the chunked reads efficiently. Using multiple reader threads keeps several such read graphs in flight concurrently, feeding the compute and writer threads continuously. 

One thread is now dedicated to the manipulation of the data. The compute thread dequeues each loaded chunk of data and calls the C++/ OpenMP kernel, which applies all transformations (restframing, centering and rebinning) in a single in-memory pass. The kernel itself is described in detail in the following section. In this case, s single compute thread is sufficient, since parallelism here comes from OpenMP within each kernel call, which already saturates all \texttt{OMP\_NUM\_THREADS} cores with \texttt{schedule(static)}. Issuing multiple concurrent kernel calls from several compute threads would oversubscribe the physical cores, degrading performance through cache contention and scheduler overhead.

Every time a compute step has been completed, the writer thread flushes the processed datasets to the output MS via \texttt{xds\_to\_table()}, the inverse \texttt{dask-ms} operation that writes a \texttt{xarray}/Dask dataset back to MS table columns chunk by chunk. A single writer is used both because CASA-format subtables are not designed for concurrent writes from multiple streams, risking race conditions and table corruption. Furthermore, the measured write throughput already saturates the available bandwidth in that direction with a single stream. This pipeline model ($\text{Read} \longrightarrow \text{Compute} \longrightarrow \text{Write}$) keeps all three threads continuously busy, effectively hiding I/O latency behind compute and viceversa.

\subsection{C++/OpenMP Kernel}

The core computation is implemented as a pybind11 extension module compiled with GCC and OpenMP. Each baseline row is processed independently, making the workload embarrassingly parallel; rows are distributed across threads to balance load while keeping synchronisation overhead low. For each row, the kernel performs the following operations, such as the restframing and centering of UVW coordinates, the phaseshift and the spectral rebinning.

The baseline vector is first scaled by $1/(1+z)$, mapping the observed coordinates to the rest-frame spatial frequency, and then rotated from the original phase centre $(\alpha_0, \delta_0)$ to the target stacking position $(\alpha_1, \delta_1)$ via a rotation matrix $R$ built from the two pointing directions:
\begin{equation}
    \begin{pmatrix} u_\mathrm{rf} \\ v_\mathrm{rf} \\ w_\mathrm{rf} \end{pmatrix}
    = R \cdot \frac{1}{1+z}
    \begin{pmatrix} u_\mathrm{obs} \\ v_\mathrm{obs} \\ w_\mathrm{obs} \end{pmatrix}.
\end{equation}

Each visibility is then multiplied by a complex fringe corresponding to the same shift of the phase centre:
\begin{equation}
    V_\mathrm{shifted}(\nu) = V_\mathrm{obs}(\nu) \cdot \exp\!\left(-2\pi i\, \frac{\nu}{c}\left(u_\mathrm{rf}\,\Delta l + v_\mathrm{rf}\,\Delta m + w_\mathrm{rf}\,\Delta n\right)\right),
\end{equation}
where $\Delta l, \Delta m, \Delta n$ are the direction cosine differences between the two phase centres. To avoid a trigonometric evaluation per channel, the phase is updated incrementally using a recurrence relation: the per-channel phase increment $\Delta\phi\cdot\Delta\nu$ is precomputed once, and subsequent channel phases are obtained by multiplying by the complex phasor $\mathrm{step} = e^{i\Delta\phi\cdot\Delta\nu}$.

The phase-shifted visibilities are projected onto the common rest-frame output grid using a channel-overlap weighting scheme identical to that of \texttt{mstransform} in frequency mode:
\begin{equation}
    V_\mathrm{out}(j) = \frac{\sum_k V_\mathrm{shifted}(k)\, w_{kj}}{\sum_k w_{kj}},
\end{equation}
with weights
\begin{equation}
    w_{kj} = \frac{\max\!\big(0,\ \min(\nu_k^+, \nu_j^+) - \max(\nu_k^-, \nu_j^-)\big)}{\Delta\nu_k},
\end{equation}
where $\nu_k^\pm$ and $\nu_j^\pm$ denote the lower and upper edges of the $k$-th input channel and the $j$-th output channel respectively, and $\Delta\nu_k$ is the input channel width, computed per channel as a centred finite difference rather than as a global median.

Visibilities are stored in \texttt{complex64} single precision in the MS, but the fringe phase involves products of baseline lengths and frequencies, whose double precision representation is essential to avoid phase errors that would partially decorrelate the stacked signal. The kernel therefore casts visibilities to \texttt{float64} before any phase multiplication and converts the result back to the original format only when writing the output.

\subsection{GPU Acceleration}
This new version of \texttt{ViSta} also allows the user to take advantage of GPU acceleration. GPU support is optional: at build time, the C++/OpenMP kernel can optionally be compiled with the \texttt{-DWITH\_CUDA} flag, otherwise the extension module will not present any CUDA dependency. When a GPU is available, the full per-row pipeline (rest-frame UVW correction, phase shift, and regridding) is offloaded to CUDA. Unlike the CPU implementation, where each MS is processed sequentially and OpenMP threads divide the $N_\mathrm{row}$ rows of a single MS among the available cores, the GPU version packs $N_\mathrm{batch}$ input MS into a single set of contiguous device buffers and launches one kernel covering all of them. One CUDA thread is assigned to each (MS, row) pair, for a total of $N_\mathrm{batch} \times N_\mathrm{row}$ threads launched concurrently, so that the geometry and frequency-grid parameters for all $N_\mathrm{batch}$ MS are uploaded together and processed in parallel on the device. The kernel call is preceded and followed by host-to-device and device-to-host memory transfers for each batch. While the GPU processes the current batch, the host continues reading and pre-processing the next one, overlapping data transfer and I/O with GPU compute.

\subsection{Post-processing}\label{sec:postprocess}

The optional post-processing operations currently still rely on the CASA-based implementation. The pre-processing and stacking stages operate on a large number of heterogeneous input datasets were the primary computational bottleneck of the pipeline, while the post-processing steps are applied to a single, already-combined output MS, where the I/O and memory pressure are substantially lower and the overhead of the CASA framework is less critical. For the final imaging step, the user can choose between \texttt{tclean} within CASA or any external imager that supports the MS format, such as WSClean \cite{Offringa2014}, which offers additional flexibility in weighting schemes and deconvolution algorithms and can be significantly faster for multi-scale imaging. Further optimisation of these stages, like the parallelisation of the continuum subtraction and the replacement of the imaging step with a direct FFT-based approach or a fully external pipeline, would be a natural next step, but falls outside the scope of this paper. Moreover, the most impactful long-term improvement would be the disposal of the MS format itself. As next-generation facilities move toward new data formats better suited for large-scale parallel processing, the remaining dependencies on CASA could be removed entirely, enabling the optimisation of the full \texttt{ViSta} pipeline.
\section{Performance Benchmarking}\label{sec:benchmarks}

The correct functioning and performance of the new \texttt{ViSta} implementation has been tested and compared to the original CASA-based pipeline. The analysis has been conducted with and without the GPU-accelerated kernel. 

\subsection{Computing Infrastructure}

The benchmarks presented in this section were carried out on Leonardo, the pre-exascale Tier-0 supercomputer of the EuroHPC Joint Undertaking (JU), hosted by CINECA and currently located at the Bologna DAMA-Technopole in Italy. We used the GPU-accelerated booster partition (\texttt{boost\_usr\_prod}), whose nodes are each equipped with 4 NVIDIA A100 GPUs (64\,GB HBM2e each) and a 32-core Intel Xeon Platinum 8358 CPU, connected via NVLink and to the rest of the system through a high-bandwidth NVIDIA Mellanox HDR100 InfiniBand network. All runs were performed on a single node with input and output MSs stored on the Lustre-based parallel filesystem.

\subsection{CASA-based pipeline comparative performance}
We have ran the original CASA-based pipeline on the same HPC system and on the same input sample as a reference. On the standard configuration described above, the programme takes a total wall-clock time of 570.1\,s (9.5\,min), broken down into 160.8\,s for rest-framing, 150.7\,s for centering, 227.6\,s for rebinning, and 31.0\,s for the final stacking step, creating three output files for each input file plus the large final one. 
\subsection{Benchmark Configurations}

We performed four scaling tests, varying one pipeline configuration parameter at a time while keeping the others fixed at a standard configuration. The input sample consists of 100 mock MSs of 200 channels of $\sim 10000$ rows, spanning a range of redshifts, telescopes and array configurations. For each run we report cumulative time spent by the reader thread(s) inside \texttt{dask.compute()} \texttt{read\_s}, the time spent by the compute thread inside the C++/OpenMP or CUDA kernel \texttt{compute\_s}, the time spent by the writer thread inside \texttt{xds\_to\_table()} \texttt{write\_s} and the total wall-clock time \texttt{wall\_s}.

An intermediated configuration has been adopted as a standard reference,consisting of $N_\mathrm{MS}=50$ MSs, $N_\mathrm{chan}=1000$ channels in the common rest-frame output spectral grid, $N_\mathrm{readers}=2$ , $\mathrm{OMP\_NUM\_THREADS}=32$ and with GPU batch size $N_\mathrm{batch}=20$ for the GPU runs. Starting from this configuration, the four tests modifies in turn one parameter at the time. For the CPU-only kernel, we varied $\mathrm{OMP\_NUM\_THREADS}$, while for the GPU kernel we varied the batch size $N_\mathrm{batch}$.

\subsection{Results}

\begin{figure}[ht]
\centering
\includegraphics[width=\textwidth]{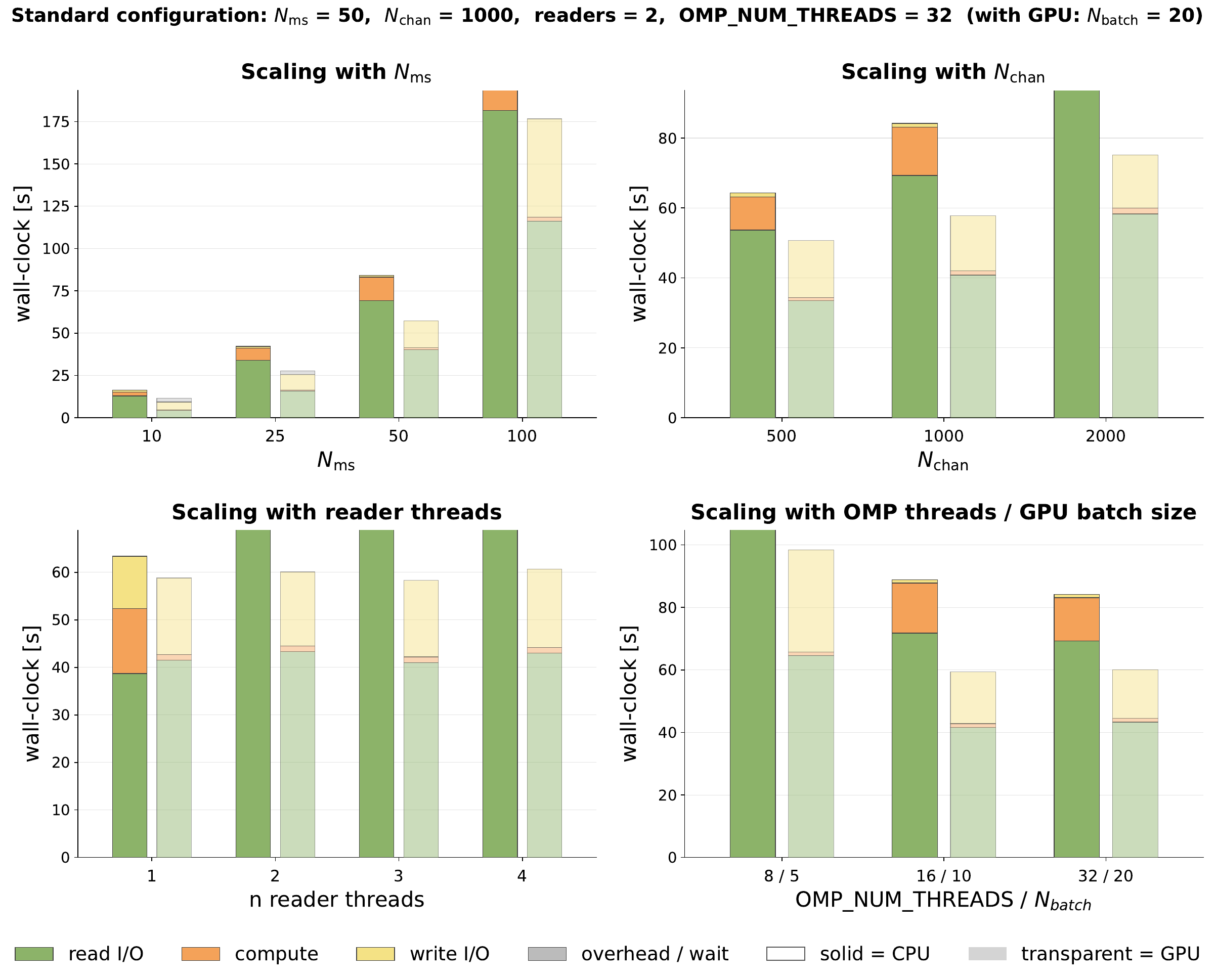}
\caption{Wall-clock time breakdown (read I/O, compute, write I/O, overhead/wait) for the CPU (\texttt{cpu\_opt}, solid bars) and GPU (A100, transparent bars) implementations, as a function of $N_\mathrm{MS}$, $N_\mathrm{chan}$, number of reader threads, and OMP\_NUM\_THREADS (CPU) / GPU batch size $N_\mathrm{batch}$ (GPU). The standard configuration is Standard configuration is: $N_\mathrm{MS}$ = 50, $N_\mathrm{chan}$ = 1000, $N_\mathrm{reader}$= 2, OMP\_NUM\_THREADS = 32 (with GPU: Nbatch = 20)}
\label{fig:benchmarks}
\end{figure}

The results of this analysis are shown in Figure~\ref{fig:benchmarks}. At the standard configuration, the CPU implementation reaches a wall-clock time of $\sim$42--43\,s, a speedup of $570.1/42 \approx 13.6\times$ over CASA. The two methods give comparable results: identical spectral resolution (25.The two methods give comparable results: identical spectral resolution (25.02 km/s, agreeing to within 0.01\%), and a mean RMS that agrees to within $\sim$1\%. Even after imaging, the noise level of the stacked cubes differs by less than 1\%, demonstrating that the physical result is fully preserved between the two implementations. Channel-to-channel differences are typical statistical fluctuations of a local RMS estimate combined with the grid offset, not a systematic bias.

\begin{table}[htpb]
\caption{Pipeline timing breakdown as a function of $N_\mathrm{MS}$.}
\label{tab:scaling_nms}
\begin{center}
\begin{tabular}{|l|l|l|l|l|l|l|l|l|}
\hline
\rule[-1ex]{0pt}{3.5ex} & \multicolumn{4}{c|}{CPU} & \multicolumn{4}{c|}{GPU} \\
\hline
\rule[-1ex]{0pt}{3.5ex} $N_\mathrm{MS}$ & read & compute & write & wall & read & compute & write & wall \\
\hline
\rule[-1ex]{0pt}{3.5ex} 10  & 12.9  & 2.3  & 1.1 & 9.4   & 4.4   & 0.2 & 4.7  & 11.6  \\
\hline
\rule[-1ex]{0pt}{3.5ex} 25  & 34.0  & 7.1  & 1.1 & 22.0  & 15.8  & 0.6 & 9.2  & 27.8  \\
\hline
\rule[-1ex]{0pt}{3.5ex} 50  & 69.3  & 13.8 & 1.1 & 43.4  & 40.2  & 1.2 & 15.9 & 50.4  \\
\hline
\rule[-1ex]{0pt}{3.5ex} 100 & 181.6 & 30.9 & 1.3 & 112.8 & 116.2 & 2.4 & 58.0 & 151.0 \\
\hline
\end{tabular}
\end{center}
\end{table}
\begin{table}[htpb]
\caption{Pipeline timing breakdown as a function of $N_\mathrm{chan}$.}
\label{tab:scaling_nchan}
\begin{center}
\begin{tabular}{|l|l|l|l|l|l|l|l|l|}
\hline
\rule[-1ex]{0pt}{3.5ex} & \multicolumn{4}{c|}{CPU} & \multicolumn{4}{c|}{GPU} \\
\hline
\rule[-1ex]{0pt}{3.5ex} $N_\mathrm{chan}$ & read & compute & write & wall & read & compute & write & wall \\
\hline
\rule[-1ex]{0pt}{3.5ex} 500  & 53.7 & 9.5  & 1.1 & 34.2 & 33.5 & 0.9 & 16.3 & 41.9 \\
\hline
\rule[-1ex]{0pt}{3.5ex} 1000 & 69.3 & 13.8 & 1.1 & 43.4 & 40.8 & 1.2 & 15.8 & 50.6 \\
\hline
\rule[-1ex]{0pt}{3.5ex} 2000 & 94.2 & 25.3 & 1.2 & 62.3 & 58.3 & 1.7 & 15.2 & 73.1 \\
\hline
\end{tabular}
\end{center}
\end{table}
\begin{table}[htpb]
\caption{Pipeline timing breakdown as a function of $N_\mathrm{readers}$.}
\label{tab:scaling_readers}
\begin{center}
\begin{tabular}{|l|l|l|l|l|l|l|l|l|}
\hline
\rule[-1ex]{0pt}{3.5ex} & \multicolumn{4}{c|}{CPU} & \multicolumn{4}{c|}{GPU} \\
\hline
\rule[-1ex]{0pt}{3.5ex} $N_\mathrm{readers}$ & read & compute & write & wall & read & compute & write & wall \\
\hline
\rule[-1ex]{0pt}{3.5ex} 1 & 38.7 & 13.7 & 11.0 & 46.6 & 41.5 & 1.2 & 16.1 & 52.2 \\
\hline
\rule[-1ex]{0pt}{3.5ex} 2 & 69.3 & 13.8 & 1.1  & 43.4 & 43.3 & 1.2 & 15.6 & 53.8 \\
\hline
\rule[-1ex]{0pt}{3.5ex} 3 & 82.6 & 17.2 & 1.3  & 43.3 & 41.0 & 1.2 & 16.1 & 51.1 \\
\hline
\rule[-1ex]{0pt}{3.5ex} 4 & 97.1 & 18.4 & 1.4  & 44.5 & 43.0 & 1.2 & 16.5 & 53.2 \\
\hline
\end{tabular}
\end{center}
\end{table}
\begin{table}[htpb]
\begin{minipage}{0.48\textwidth}
\centering
\caption{Pipeline timing breakdown as a function of OMP\_NUM\_THREADS (CPU).}
\label{tab:scaling_omp}
\vspace{2mm}
\begin{tabular}{|l|l|l|l|l|}
\hline
\rule[-1ex]{0pt}{3.5ex} OMP & read & compute & write & wall \\
\hline
\rule[-1ex]{0pt}{3.5ex} 8  & 121.7 & 20.1 & 6.9 & 81.8 \\
\hline
\rule[-1ex]{0pt}{3.5ex} 16 & 71.8  & 16.0 & 1.1 & 43.3 \\
\hline
\rule[-1ex]{0pt}{3.5ex} 32 & 69.3  & 13.8 & 1.1 & 43.4 \\
\hline
\end{tabular}
\end{minipage}
\hfill
\begin{minipage}{0.48\textwidth}
\centering
\caption{Pipeline timing breakdown as a function of $N_\mathrm{batch}$ (GPU).}
\label{tab:scaling_batch}
\vspace{2mm}
\begin{tabular}{|l|l|l|l|l|}
\hline
\rule[-1ex]{0pt}{3.5ex} $N_\mathrm{batch}$ & read & compute & write & wall \\
\hline
\rule[-1ex]{0pt}{3.5ex} 5  & 64.6 & 1.1 & 32.7 & 76.3 \\
\hline
\rule[-1ex]{0pt}{3.5ex} 10 & 41.6 & 1.2 & 16.6 & 51.9 \\
\hline
\rule[-1ex]{0pt}{3.5ex} 20 & 43.3 & 1.2 & 15.6 & 53.8 \\
\hline
\end{tabular}
\end{minipage}
\end{table}

\paragraph{Scaling with $\mathbf{N_\mathrm{MS}}$, Tab. \ref{tab:scaling_nms}.} On both CPU and GPU, \texttt{read\_s} and \texttt{wall\_s} scale close to linearly with $N_\mathrm{MS}$, since both the total data volume and the number of per-MS kernel calls grow linearly with the sample size. The GPU \texttt{write\_s}, however, grows super-linearly between $N_\mathrm{MS}=50$ and $100$.  Indeed the GPU processes batches of 20 MSs almost instantly, so the writer queue backs up and the run becomes writer-bound. 

\paragraph{Scaling with $\mathbf{N_\mathrm{chan}}$,
Tab. \ref{tab:scaling_nchan}.} CPU \texttt{compute\_s} scales close to linearly with $N_\mathrm{chan}$, consistent with the rebinning function being an $O(N_\mathrm{chan})$ operation per row. \texttt{read\_s} also increases with $N_\mathrm{chan}$, since the output buffers scale with the output grid size. On GPU, \texttt{compute\_s} barely changes because the rebinning loop is fully parallel across output channels on the device. Finally, \texttt{write\_s} stays roughly constant in this regime, depending mainly on the number of rows rather than on $N_\mathrm{chan}$.

\paragraph{Scaling with the $\mathbf{N_\mathrm{reader}}$, Tab. \ref{tab:scaling_readers}.} On CPU, with a single reader the writer-side contribution to the wall time is anomalously high, because the writer thread repeatedly stalls waiting for the lone reader to keep up, draining the pipeline serially at the end of the run. Moving to $N_\mathrm{readers}=2$ removes this stall almost entirely and gives the best overall wall time (43.4\,s). Beyond 2 readers, \texttt{read\_s} keeps increasing but \texttt{wall\_s} stays essentially flat: once compute and writer are no longer starved, extra readers add filesystem contention without reducing the wall-clock time. On GPU instead, the bottleneck is the single writer thread, so feeding the compute stage faster via more readers has no effect on the overall time and the pipeline is writer-bound regardless of how quickly data arrives.

\paragraph{Scaling with OMP\_NUM\_THREADS or GPU batch size, Tab. \ref{tab:scaling_omp}. and \ref{tab:scaling_batch}.} These two parameters control different things and are shown together on a shared axis only for visual comparison. \texttt{OMP\_NUM\_THREADS} sets the fine-grained parallelism of the C++ kernel itself, distributing baseline rows across cores. The GPU batch size $N_\mathrm{batch}$ instead does not affect the GPU's internal parallelism but sets how many MS are grouped into a single host-device kernel launch, controlling the granularity of pipeline overlap between reading, GPU compute, and writing. For the CPU, \texttt{compute\_s} decreases with the number of OMP threads, but with strongly diminishing returns: doubling the thread count from 16 to 32 yields only a marginal reduction, far less than the speedup one might naively expect. This is consistent with the per-row kernel being memory-bandwidth bound rather than compute bound at high thread counts. The much larger improvement when going from OMP=8 to OMP=16 is also reflected in \texttt{read\_s} and \texttt{write\_s}, both of which drop substantially over the same step: with too few OMP threads the compute thread is slow enough that the reader and writer threads queue up waiting on it, inflating their measured times. On the other hand, GPU \texttt{compute\_s} step is essentially insensitive to batch size. Thee kernel-launch overhead is amortised regardless of how many MS are packed together, but at the sample size and dataset dimensions used here the GPU is so far from saturation that any such effect remains too small to be visible in these measurements. \texttt{write\_s}, however, is highest at the smallest batch size tested and drops substantially for larger batches: with very small batches the GPU returns results frequently, but each batch is too small to overlap efficiently with the surrounding read/write cycle, so per-batch overhead dominates. Intermediate batch sizes are therefore close to the sweet spot, balancing kernel-launch overhead against pipeline overlap. Since the GPU run is writer-bound rather than compute-bound (Figure~\ref{fig:benchmarks}, bottom-right panel), increasing OMP\_NUM\_THREADS in the GPU configuration would be expected to reduce \texttt{read\_s} only marginally, with no measurable effect on \texttt{wall\_s}. For this reason it was not varied independently in the GPU runs.

These tests demonstrate that the new pipeline achieves substantial speedups over the original CASA-based implementation, with the CPU configuration comfortably I/O-bound and the GPU configuration writer-bound, while compute plays only a secondary role except at very low thread counts. The CPU-only configuration is already highly efficient for the data volumes considered here, and the GPU path is expected to become increasingly advantageous for much larger samples, where the compute savings accumulate, or on local workstations with fewer CPU cores where the OpenMP kernel cannot reach the same throughput. Further optimization on the output side is constrained by the MS format, which does not support concurrent writes from multiple streams. Relieving the writer bottleneck would therefore require redesigning the output format, which we leave for future updates of the code.

\section{Conclusions}\label{sec:conclusions}

We have presented an HPC-optimized reimplementation of \texttt{ViSta}, the new stacking methodology that combines interferometric datasets from sources at different redshifts, observed with different telescopes and configurations, directly in the Fourier domain. In T25 we had already demonstrated the scientific advantages of working in the visibility plane. However, the main bottleneck preventing these advantages from being fully exploited was that the original implementation was memory and computationally demanding. The new implementation replaces the original CASA-based pipeline with a self-contained Python package backed by a C++/OpenMP core, with optional GPU acceleration.

\begin{figure}[ht]
\centering
\includegraphics[width=\textwidth,trim={3.2cm 0 0 0}]{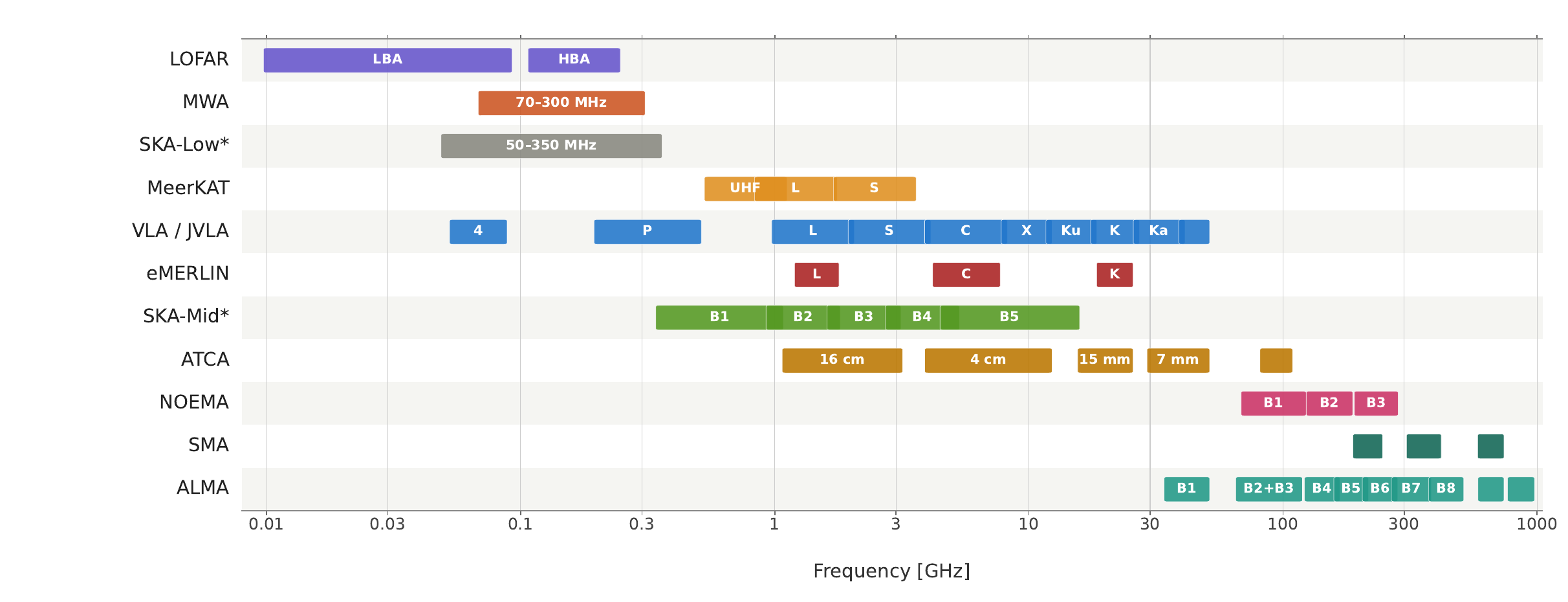}
\caption{Frequency coverage of the major radio and sub-millimetre interferometric arrays, from the low-frequency regime  to the sub-millimetre. The horizontal axis shows frequency in GHz on a logarithmic scale. Each coloured bar represents the nominal frequency range of a receiver band.}
\label{fig:interferometers}
\end{figure}

We have tested the performance of his new \texttt{ViSta} version, and we have shown that even the CPU-only configuration already exceeds the original CASA-based pipeline by more than an order of magnitude. Should the need arise to combine increasingly large numbers of MSs, the GPU-accelerated kernel allows the pipeline to retain this performance and keep execution times under control as the sample size grows.
The code repository is publicly hosted \href{https://github.com/martitors/ViSta-HPC}{here}. This reimplementation makes \texttt{ViSta} practical as a general purpose archival tool. As we have seen, since each input dataset is independently rescaled, re-centred, and regridded onto a common rest-frame \textit{uv}-plane before being combined, the method naturally allows to combining observations from different interferometers, frequencies, and epochs regardless of array configuration or spectral setup. 
While the original \texttt{ViSta} was validated exclusively on simulated and real ALMA data, the new implementation is fully general and it is immediately applicable to any interferometric facility that adopts this standard, including the VLA, MeerKAT, eMERLIN, NOEMA, and others. Figure~\ref{fig:interferometers} illustrates the frequency coverage of the major radio and sub-millimetre arrays currently in operation. When observing objects belonging to the same class but at different redshifts, on which we aim to perform the stacking analysis, their emission is shifted across the frequency axis and the overlap between the bands of different telescopes increases substantially. 
This makes it possible to exploit data across facilities that would otherwise remain unused in interferometric archives. Indeed, the majority of archival visibilities contain information beyond the original science goal, such as spectral lines or sources that were not the target of the observation may still be embedded in the data, sitting unused while resources continue to be spent to preserve them. Stacking the raw data provides a way to recover this otherwise dormant information, further increasing the legacy value of the archive. Finally, as next-generation facilities move toward new data formats, \texttt{ViSta} will be able to adapt accordingly, eventually enabling a complete pipeline processing without any dependency on CASA-based routines.

Nowdays, the ability of \texttt{ViSta} of retrieving information in the Fourier domain is particularly relevant in view of the SKAO era. \texttt{ViSta} can provide a natural framework for quantifying how much information is lost when moving from the \textit{uv}-plane to the image plane, which is an increasingly important question for future surveys where visibilities will not be routinely preserved and most data products will only be available as images. In this sense, \texttt{ViSta} serves both as a tool for maximizing the scientific return of current and archival interferometric data, and as a benchmark for assessing what will and will not be recoverable in the image-dominated data products of future advanced facilities.
\section*{Acknowledgments}
All authors contributed equally to the planning of the reimplementation, to the discussion of the results, and to the writing of the present manuscript. M.T. is responsible for the development of the code and performed the described analysis and benchmarks. The benchmarks were carried out on the Leonardo supercomputer at CINECA. The new pipeline makes use of the following software: Python \cite{Python3}, NumPy \cite{numpy}, dask-ms \cite{daskms}, Dask \cite{dask}, xarray \cite{xarray}, pybind11 \cite{pybind11}, OpenMP \cite{openmp}, CUDA \cite{cuda}, casacore \cite{casacore}, CASA \cite{CASATeam2022}, WSClean \cite{Offringa2014}, UVMultiFit \cite{MartiVidal2014}.

\bibliography{report} 
\bibliographystyle{spiebib} 

\end{document}